\documentclass{revtex4}
  \usepackage{graphicx}
 \usepackage{amssymb, amsmath,amsxtra}
 \usepackage{bm}
 \usepackage{epsfig,subfigure}
 \usepackage{colordvi}
 \usepackage{color}
 \usepackage{pstricks,pst-node,pst-text,pst-3d}

\begin{document}

\title{\textbf{Nonlinear Acoustic Waves in Channels with Variable Cross Sections}}
\author{Vladimir F.~Kovalev \\
 \textit{Keldysh Institute of Applied Mathematics, Russian Academy of Sciences, } \\
   \textit{Moscow, 125047 Russia} \\
   and \\
 Oleg V.~Rudenko \\
\textit{Faculty of Physics, Moscow State University, Moscow, 119991 Russia} }

\begin{abstract}
The point symmetry group is studied for the generalized Webster-type equation describing non-linear acoustic waves in lossy channels with variable cross sections. It is shown that, for certain types of cross section profiles, the admitted   symmetry group is extended and the invariant solutions corresponding to these profiles are obtained. Approximate analytic solutions to the generalized Webster equation are derived for channels with smoothly varying cross sections and arbitrary initial conditions.
\end{abstract}

\maketitle

\section{Introduction}

The Webster equation \cite{web-pnac-1919,eisner-bk-1964,lan-bk-1986} describes waves propagating in pipes, horns, concentrators, and other waveguides characterized by a varying cross section $S(x)$:
\begin{equation}
 \label{intr1}
 \frac{1}{S(x)}\, \frac{\partial}{\partial x} \left( S(x) \frac{\partial p}{\partial x} \right)
 - \frac{1}{c^2}\, \frac{\partial^2 p}{\partial t^2}  = 0 \,,
\end{equation}
where $x$ is the coordinate measured along the waveguide axis. Equation (\ref{intr1}) is applicable to pipes
whose characteristic width is small compared to the wavelength. In addition, the cross section is assumed
to vary slowly along the $x$-axis: the area $S(x)$ varies only slightly as $x$ varies by a quantity on the order of the pipe width \cite{lan-bk-1986}. Note that, in the general case, Eq. (\ref{intr1}) cannot describe a wave propagating in the pipe in one direction (e.g., in the positive direction of the $x$ axis):
\begin{equation}
 \label{intr2}
 p(x,t) = A(x) \Phi \left( t-\varphi (x) \right)\,,
\end{equation}
where $\Phi$ is an arbitrary function. Indeed, substituting Eq. (\ref{intr2}) in Eq. (\ref{intr1}), we obtain the relations
\begin{equation}
 \label{intr3}
 \left( \frac{{\rm d}\varphi }{{\rm d} x} \right)^2 = \frac{1}{c^2}\,, \quad  \frac{{\rm d} }{{\rm d} x} \left( A^2 S \right) = 0  \,, \quad  \frac{{\rm d}  }{{\rm d} x}
 \left( S\, \frac{{\rm d} A }{{\rm d} x} \right) = 0 \,.
\end{equation}
For the boundary conditions $A(x=0)=p_0$, $S(x=0)=1$,
we obtain from Eqs. (\ref{intr3})
\begin{equation}
 \label{intr4}
 A= \frac{p_0 }{1 \pm x/R }\,, \quad  S=\left( 1 \pm x/R \right)^2 \,.
\end{equation}

Here, the plus sign corresponds to a spherically expanding pipe and, hence, a spherically divergent
wave. If, in Eqs. (\ref{intr4}), we take a minus sign, we obtain a tapered pipe and a wave converging to the center of the sphere $x=R$. As $R\to \infty$, the pipe becomes homogeneous ($S(x)=const$) and the wave becomes plane. \par

Thus, it is only in the case of a plane or spherical wave that (\ref{intr1}) describes a wave traveling in one direction with an arbitrary profile $\Phi(t)$. In all other cases, the inhomogeneity of the cross section $S(x)$ causes multiple reflections of signals propagating in opposite directions, which leads to the formation of a complicated wave field pattern as a combination of standing and traveling waves. Study of the influence of inhomogeneity $S(x)$ on the field structure gives a chance to solve the inverse problem, i.e. it helps to reconstruct the unknown function $S(x)$ using characteristics of the reflected (transmitted) wave. This problem was discussed in \cite{rud-acp-2010} for small-amplitude acoustic waves in a tapered waveguide and also in \cite{rud-acj-1990} for finite-amplitude waves. Solution to the inverse problem can serve as a classical illustration of the effect of the dimensional reduction that was discussed recently in \cite{sh-pepan-2010,shf-tmp-2011,shf-jpa-2012}.
 \par

However, in the case of slowly varying S(x), within distances on the order of several wavelengths, it is possible to neglect multiple reflections and consider a wave traveling in one direction. This statement holds
for both conventional Webster equation (1) and its generalization to the case of a nonlinear dissipative
medium filling the pipe. The generalized Webster-type equation appears in problems on propagation of
intense sound \cite{rud-ufn-1995,rud-aaa-2002}. It differs from Eq. (\ref{intr1}) in that it contains two additional terms describing nonlinear and dissipative effects. We represent this equation in the form
\begin{equation}
 \label{intrweq5}
  \frac{\partial^2 p}{\partial t^2}  - c^2  \frac{\partial^2 p}{\partial x^2} = c^2 \frac{\partial \ln S(x)}{\partial x}  \frac{\partial p}{\partial x} +
 \frac{\varepsilon}{c^2 \rho} \frac{\partial^2 p^2}{\partial t^2}
 + \frac{b}{\rho} \frac{\partial^3 p}{\partial x \partial t^2}  \, .
\end{equation}
Here, $\varepsilon$ and $b$ are the nonlinearity and dissipation parameters (the notations are the same as those given
in \cite{rud-bk-1997}) and $\rho$ is the density of the medium. Both the initial (\ref{intr1}) and generalized (\ref{intrweq5}) Webster equations are suitable for describing traveling and standing waves.
 \par

In the situation where each of the terms appearing on the right-hand side of the equation is small compared to the terms appearing on the left-hand side, a traveling wave can be considered. In this case, using the method of a slowly varying profile \cite{rud-bk-1997}, it is possible to reduce the order of nonlinear equation (\ref{intrweq5}). Following the standard procedure \cite{rud-bk-1990}, we change from the variables $x$ and $t$ involved in Eq. (\ref{intrweq5}) to new independent variables: the "slow" coordinate $x_1=\delta x$ (where $\delta$ is the small parameter of the problem) and the time $\tau = t-x/c$ in the coordinate system traveling with the velocity of sound. Ignoring small terms on the order of $\delta^n$, where $n \geqslant 2$, we arrive at the evolution equation
\begin{equation}
 \label{intrweq6}
 \frac{\partial p}{\partial x} -
 \frac{\varepsilon}{c^3 \rho} p \frac{\partial p}{\partial {\tau}} - \frac{b}{2 c^3 \rho}
  \frac{\partial^2 p}{\partial {\tau}^2} + \frac{p}{2} \frac{\partial}{\partial x}  \left( \ln S(x) \right)  = 0 \, .
\end{equation}
This equation is used not only as a model of wave propagation in a pipe but also for calculating the acoustic field in an inhomogeneous medium in the geometric acoustics approximation \cite{rud-ufn-1995,rud-acj-1994}; in the latter case, it plays the role of the transfer equation represented in terms of the ray coordinates. The axis of a ray tube is the geometric ray calculated from the eikonal equation, and the function $S(x)$ represents the cross section of the ray tube. \par

The study of even the simplest linear Webster equation (\ref{intr1}) reveals rather interesting effects, such as, e.g.,
tunneling of sound waves in the case of sound propagation through a tapered waveguide \cite{rud-acp-2010}. The change to generalized Webster equation (\ref{intrweq6}) (GWE), which contains additional contributions due to nonlinear effects and absorption, makes it possible to study problems on the propagation of finite-amplitude sound waves in absorbing media and, in particular, acoustic monitoring of propagation media \cite{rud-acj-1990}. For these purposes, exact and approximate analytic solutions to the finite-amplitude acoustic wave equation acquire special importance, as in the case of the Burgers equation and its generalizations for homogeneous media (see \cite{rud-aaa-2002} and Sect. 2, Ch. 7 in [11]). \par

Although solutions to the GWE simultaneously allowing for the effects of nonlinearity, absorption, and inhomogeneity are important for describing the behavior of sound waves in nonlinear absorbing media, their analytic derivation is a difficult problem even in the case of using approximate methods. The standard practice is either to neglect dissipation, which makes it possible, by changing the independent variable, to reduce the initial equation to the Hopf equation \cite{rud-aaa-2002}, or to assume that the nonlinearity is small and to solve the sound wave equation by the method of successive approximations; the latter approach has been used to analyze the second harmonic behavior in a sound channel with a variable cross section \cite{rud-acj-1990}. On the other hand, it is evident that, as the cross section of the channel decreases, the second harmonic amplitude grows faster than the fundamental harmonic amplitude, which necessitates analyzing the GWE for a finite-amplitude sound wave. This paper is devoted to finding an analytic solution to the GWE under these conditions. \par

The paper contains five sections. The second section formulates the initial equations for the theoretical analysis of sound wave propagation in a medium with allowance for nonlinearity and absorption. In addition to the initial GWE, we consider its modified analog, which simplifies the analytic investigation by group theoretical methods. The latter prove to be the most effective instrument for constructing analogs of the exact analytic solutions obtained earlier for a homogeneous medium. In the third section, we determine the point transformation group for the modified GWE and show that, for specific cross section profiles, this symmetry group can be extended. We determine the invariant solutions corresponding to the aforementioned symmetries and compare these solutions with those obtained for a waveguide with a constant cross section. The fourth section is devoted to constructing an approximate analytic solution to the GWE for arbitrary varying cross sections and arbitrary initial conditions. The consideration is based on the theory of approximate transformation groups, which allows determination of an approximate symmetry and the corresponding approximate analytic solution. The small parameter used in the calculations is the slowness of the variation in the waveguide cross section profile $S(x)$. In the fifth section, we briefly formulate the main results of our study.

\section{Basic equations \label{equations}}

To analyze the nonlinear effects that accompany sound wave propagation in an absorbing medium in a channel with a variable cross section $S(x)$, we use GWE (\ref{intrweq6}), which can be represented in the form
\begin{equation}
 \label{webeq1}
 \frac{\partial p}{\partial x} - a p \frac{\partial p}{\partial \tau} -\nu \frac{\partial^2 p}{\partial \tau^2} + \frac{p}{2} \frac{\partial }{\partial x} \left( \ln S(x) \right) = 0 \,, \quad
 p(0,\tau)=P(\tau)\, .
\end{equation}
Unlike the equations discussed above, Eq. (\ref{webeq1}) is expressed in dimensionless form. The change from the
physical variables involved in Eq. (\ref{intrweq6}) to the more convenient normalized variables appearing in Eq. (\ref{webeq1}) is performed through the following substitution in Eq. (\ref{intrweq6}):
 \[
 x\rightarrow \frac{c}{\omega} \, x \,, \quad  \tau \rightarrow \frac{\tau}{\omega} \,, \quad p \rightarrow p_0 p \,.
  \]
Here, the normalizing constants $\omega$ and $p_0$ have the meaning of the characteristic frequency and signal
amplitude values, respectively. The two parameters involved in Eq. (\ref{webeq1}) are determined by the following
dimensionless combinations of constants:
 \[
  a=\frac{\varepsilon p_0}{c^2 \rho} \,,  \quad \nu = \frac{b \omega}{2 c^2 \rho} \,.
 \]
Their ratio $a/\nu$ is called the acoustic Reynolds number \cite{rud-bk-1997}. It characterizes the relative contributions of nonlinear and dissipative effects to the distortion of the wave profile. When $a/\nu$ is large, nonlinear effects predominate; when this quantity is small, dissipative effects are dominant. Without loss of generality, in Eq. (\ref{webeq1}) we set $S(0)=1$.
\par

We can eliminate the last term from Eq. (\ref{webeq1}) by changing the variable $x$ and introducing the absorption
as a function $\mu$ of the coordinate along the channel:
\begin{equation}
  \label{zamena}
 \zeta = \int {\rm d}x / \sqrt{S(x)}\,, \quad p \sqrt{S} = u \,, \quad \mu = \nu \sqrt{S(x(\zeta))}\,,
\end{equation}
Then, in terms of the new variables, Eq. (\ref{webeq1}) takes the form
\begin{equation}
 \label{webeq1a}
  \frac{\partial u}{\partial \zeta} - a u \frac{\partial u}{\partial \tau}
  - \mu \frac{\partial^2 u}{\partial \tau^2}  = 0 \,, \quad
 u(0,\tau)=P(\tau)\, .
\end{equation}
Introducing the new variable $q$ related to $u$ by the formula $u = 2 (\partial q/\partial \tau) $, we replace Eq.~(\ref{webeq1}) by the modified GWE (MGWE)
\begin{equation}
 \label{webeq2}
  \frac{\partial q}{\partial \zeta} - a  \left( \frac{\partial q}{\partial \tau} \right)^2
  -\mu  \frac{\partial^2 q}{\partial \tau^2} = 0 \,, \quad
  q(0,\tau)=W(\tau)\, .
\end{equation}
The change to the new variable $q$ in Eq.~(\ref{webeq1a}) increases the order of this equation. However, its single integration with respect to $\tau$ yields evolutional equation (\ref{webeq2}) for $q$. This procedure determines $q$ accurate to the function $C(\zeta)$ (in Eq.~(\ref{webeq2}), this function is omitted), whose choice is fairly arbitrary. For example, for solutions to Eq.~(\ref{webeq2}) that are periodic in $\tau$, this function can be chosen so as to make the period average value of $q$ zero for any values of $\zeta$. At the same time, it is evident that the choice of $C(\zeta)$ does not affect the physical meaning of $u$. \par

\section{Symmetry group and invariant solutions to the generalized Webster equation \label{symmetry}}

For Eq.~(\ref{webeq2}) in the case of an arbitrary inhomogeneity profile $\mu(\zeta)$, the admitted point transformation
group is given by three infinitesimal operators:
 \begin{equation}
  \begin{aligned}
  X_1 = \frac{\partial }{\partial{\tau}}  \,, \quad
  X_2 = \frac{\partial }{\partial{q}}  \,, \quad
  X_3 = \zeta \frac{\partial }{\partial{\tau}}  - \frac{\tau}{2a} \frac{\partial }{\partial{q}}  \,.
 \label{genweb2}
 \end{aligned}
 \end{equation}
The first two of them represent the translation operators with respect to the variables $\tau$ and $q$, which are evident from the physical point of view; the third operator corresponds to the Galilean transformation group. Transformation group (\ref{genweb2}) can be extended for the cross section profiles of a specific type:
\begin{equation}
 \label{classify}
 M \left( \frac{ {\rm d} }{{\rm d} \zeta} \ln (\mu (\zeta) ) \right)^{-1} = b(\zeta)\,, \quad b(\zeta) = \beta_0 + \beta_1 \zeta + \beta_2 \zeta^2 \,, \quad M = {\rm const} \neq 0 \,.
\end{equation}
Condition (\ref{classify}) plays the role of the classifying relation separating the specific profile types for which transformation group (\ref{genweb2}) is extended by the additional operator $X_4$:
 \begin{equation}
  \begin{aligned}
  X_4 = b \frac{\partial }{\partial {\zeta}}  + \frac{\tau}{2} \left( M + \frac{{\rm d}b}{{\rm d}\zeta} \right) \frac{\partial }{\partial {\tau}} + \left( M q - \frac{(\tau^2+2 \int{\rm d} \zeta \mu)}{8a}
   \frac{{\rm d}^2 b}{{\rm d}\zeta^2} \right) \frac{\partial }{\partial {q}} \,.
 \label{genweb2ext1}
 \end{aligned}
 \end{equation}
Classifying relation (\ref{classify}) is a first-order differential equation for the function $\mu(\zeta)$; being explicitly integrated, this equation determines a three-parameter (the parameters are $\beta_0$, $\beta_1$, and $\beta_2$) family of curves in the $\{\zeta, \mu\}$ space:
\begin{equation}
 \label{classify2}
  \ln (\mu /\nu ) = d(\zeta)\equiv M \int\limits_{0}^{\zeta} \frac{{\rm d} y}{b(y)} \,,
  \end{equation}
where the form of the function $d(\zeta)$ depends on the relative values of the parameters $\beta_i$:
\begin{equation}
 \label{classify3}
  d(\zeta)=
 \begin{cases} \frac{2 M}{\sqrt{4\beta_0\beta_2-\beta_1^2}} \left[ \arctan\frac{\beta_1+2\beta_2 \zeta}{\sqrt{4\beta_0\beta_2-\beta_1^2}} - \arctan \frac{\beta_1}{\sqrt{4\beta_0\beta_2-\beta_1^2}} \right] \,, & \quad \beta_1^2 < 4\beta_0\beta_2 \,, \\
  \frac{M}{\sqrt{\beta_0\beta_2}}\frac{z}{z+\sqrt{\beta_0/\beta_2}}  \,, & \quad \beta_1^2 = 4\beta_0\beta_2 \,, \\
  \frac{M}{\sqrt{\beta_1^2-4\beta_0\beta_2}} \ln\frac{(\sqrt{\beta_1^2-4\beta_0\beta_2} -\beta_1-2\beta_2 \zeta)(\sqrt{\beta_1^2-4\beta_0\beta_2} +\beta_1)}{(\sqrt{\beta_1^2-4\beta_0\beta_2} +\beta_1+2\beta_2 \zeta)(\sqrt{\beta_1^2-4\beta_0\beta_2} -\beta_1)} \,, & \quad \beta_1^2 > 4\beta_0\beta_2 \,.
  \end{cases}
\end{equation}
The choice of $M =0$ corresponds to a channel with a constant cross section, ${{\rm d} \mu (\zeta)}/{\rm d} \zeta = 0$; then, classifying relation (\ref{classify}) is automatically satisfied for any $\beta_i$. In this case, the following three operators appear instead of the operator $X_4$:
 \begin{equation}
  \begin{aligned}
  X_{41} = \frac{\partial }{\partial{\zeta}}  \,, \quad
  X_{42} = \zeta \frac{\partial }{\partial{\zeta}} + \frac{\tau}{2} \frac{\partial }{\partial{\tau}} \,, \quad
  X_{43} = \zeta^2 \frac{\partial }{\partial{\zeta}}  + \tau \zeta \frac{\partial }{\partial{\tau}}
  - \frac{(\tau^2+2\zeta \nu)}{4a} \frac{\partial }{\partial{q}}  \,.
 \label{genweb2ext2}
 \end{aligned}
 \end{equation}
The first of them, $X_{41}$, is the translation operator along the $\zeta$ axis; the second operator, $X_{42}$, represents the dilation transformation; and $X_{43}$ corresponds to the projective transformation group. In addition to operators (\ref{genweb2ext2}), for a channel with a constant cross section, $\mu \equiv \nu$, the MGWE also allows the infinite subgroup operator
 \begin{equation}
  \begin{aligned}
  X_{\infty} = k(\zeta,\tau) \exp \left( - \frac{a q}{\mu} \right) \frac{\partial }{\partial{q}} \,, \qquad \frac{\partial k}{\partial{\zeta}}
  - \mu \frac{\partial^2 k}{\partial{\tau^2}} = 0 \,.
 \label{genweb2ext3}
 \end{aligned}
 \end{equation}
Here, the linear parabolic equation, which is satisfied by the function of two variables $k(\zeta,\tau)$, can be represented in terms of the variables $\{x,\tau\}$:
 \[ \frac{\partial k}{\partial{x}}
  - \nu \frac{\partial^2 k}{\partial{\tau^2}} = 0 \,.
 \]
The latter fact will be used by us in constructing the approximate point symmetry for the MGWE. We note that the symmetry group given by Eqs. (\ref{genweb2}), (\ref{genweb2ext2}), and (\ref{genweb2ext3}) is well known in the theory of the modified Burgers equation \cite{ibr-spr-1994}, to which the MGWE is reduced in the case under consideration.  \par

In constructing the invariant solutions to the MGWE, we concentrate on studying solutions that are invariant with respect to the one-parameter group with the operator $X_4$, because this operator represents an analog of linear combinations of operators (\ref{genweb2ext2}), the use of which for channels with constant cross sections yields the well-known and physically meaningful particular solutions to the nonlinear Burgers equation. \par

Taking into account the initial conditions for the MGWE (\ref{webeq2}), we represent the desired solutions by using two invariants $J_1$ and $J_2$ and the operator $X_4$, where
 \begin{equation}
  \begin{aligned} &
  J_{1} = q {\rm e}^{- d}  + \frac{\beta_2}{2a} \left[ \frac{\zeta \tau^2}{2 b} {\rm e}^{- d}
         + \int\limits_{0}^{\zeta} \frac{{\rm d} \zeta'}{b(\zeta')} {\rm e}^{- d(\zeta')}
         \int\limits_{0}^{\zeta'} {\rm d} \zeta'' {\rm e}^{d(\zeta'')} \right] \,, \quad 
         J_{2} = \frac{\tau}{\sqrt{b}} {\rm e}^{- d/2} \,,
 \label{invariants}
 \end{aligned}
 \end{equation}
as $J_1=W(J_2)$ or, in an explicit form, as
 \begin{equation}
  \begin{aligned}
  q = & {\rm e}^{d(\zeta)} \left\{W(\lambda) - \frac{\beta_2}{2a} \left[ \frac{\zeta \tau^2}{2 b} {\rm e}^{- d(\zeta)}
         + \int\limits_{0}^{\zeta} \frac{{\rm d} \zeta'}{b(\zeta')} {\rm e}^{- d(\zeta')} \int\limits_{0}^{\zeta'} {\rm d} \zeta'' {\rm e}^{ d(\zeta'')} \right] \right\}\,,  \ 
         \lambda = \frac{\tau {\rm e}^{- d/2}}{\sqrt{b}}  \,.
 \label{solutiongen}
 \end{aligned}
 \end{equation}
To solve the MGWE, we substitute the latter expression in the initial equation, then we obtain a second-order ordinary differential equation for the function $W(\lambda)$:
 \begin{equation}
  \begin{aligned}
 \frac{{\rm d}^2 W}{{\rm d} {\lambda^2}} + a \left(\frac{{\rm d} W}{{\rm d} {\lambda}}\right)^2 + \left( M + \beta_1 \right) \frac{\lambda}{2} \left(\frac{{\rm d} W}{{\rm d} {\lambda}}\right) - M W  + \frac{\beta_0 \beta_2}{4a} {\lambda}^2 = 0 \,.
 \label{factoreq}
 \end{aligned}
 \end{equation}
As examples, we consider several specific cases of the general relations.
\par

\textit{Example 1:} $\beta_2=0$, $\beta_0, \beta_1\neq 0$.
\par

This type of a channel with a constant cross section corresponds to a combination of the translation operator along the $x$ axis and the dilation group. For $\beta_2=0$, from Eqs. (\ref{classify3}) and (\ref{factoreq}) it follows that the solution is self-similar (compare with \cite{rud-aaa-2002}); depending on the relative values of the parameters $\beta_1$ and $M$, it is realized for the following functions $S(x)$:
\begin{equation}
 \label{profile}
  S(x)=
 \begin{cases} \left( 1 + \left(M+\beta_1 \right) x/\beta_0 \right)^{2M/(\beta_1+M)} \,, \quad \beta_1/M \neq -1 \,,  \\
  \exp (2Mx/\beta_0)\,, \quad \beta_1/M = -1  \,.
  \end{cases}
\end{equation}
For an exponential dependence of the cross-sectional area on $x$, i.e., for $\beta_1/M =-1$, the variable $\lambda=\tau/\sqrt{\beta_0}$ does not depend on $\zeta$; then, the solution to Eq. (\ref{factoreq}) is expressed in integral form
\begin{equation}
 \label{solution1}
  \lambda = \int {\rm d} W \left( C_0 \exp{(-2aW)} + (M/2a^2)(2aW-1) \right)^{-1/2} + C_1 \,,
  \quad C_0, C_1 = {\rm {const}} \,,
\end{equation}
and has bounded periodic solutions for $M<0$ and $C_0 <0 $.

For a power-law dependence of the cross-sectional area on $x$, the self-similar variable $\lambda$ and the function $d$
appearing in Eq. (\ref{solutiongen}) have the form
\begin{equation}
 \label{selfsimilar}
 \lambda = \frac{\tau}{\sqrt{\beta_0}} \left( 1 + \frac{\beta_1+M}{\beta_0} x \right)^{\frac{\beta_1-M}{2(\beta_1+M)}} \,,
 \quad d = \frac{M}{\beta_1} \ln \left( 1+ \frac{\beta_1}{\beta_0} \zeta \right)\,,
 \quad \beta_1 + M \neq 0  \,.
\end{equation}

\textit{Example 2:} $\beta_1=0$, $\beta_0, \beta_2 \neq 0$. \par

This type of a channel with a constant cross section corresponds to a combination of translation operators
along the $x$ axis and the projective transformation group. In our case of a channel with a varying cross
section, the functions $\lambda$ and $d$ appearing in Eq. (\ref{solutiongen}) take the form
\begin{equation}
 \label{selfsimilar2}
 \lambda = \frac{\tau {\rm e}^{-d/2}}{\sqrt{\beta_0+\beta_2 \zeta^2}} \,,
 \quad d = \frac{M}{\sqrt{\beta_0 \beta_2}} \arctan \left( \sqrt{\frac{ \beta_2}{\beta_0}}\zeta \right)\,,
 \quad \beta_1 = 0  \,,
\end{equation}
where the function $\zeta$ is related to $x$ by the formulas that implicitly determine the dependence of the cross-sectional area $S$ on $x$:
\begin{equation}
 \label{selfsimilar3}
 \zeta = \sqrt{\frac{ \beta_0}{\beta_2}} \tan \left( \frac{\sqrt{\beta_0 \beta_2}}{2M} \ln S  \right)  \,,
 \quad x= \frac{\beta_0}{2M} \int\limits_{1}^{S} \frac{{\rm d} S}{\sqrt{S}}
 \cos^{-2} \left( \frac{\sqrt{\beta_0 \beta_2}}{2M} \ln S  \right) \,.
\end{equation}
We note that the self-similar solutions obtained for the exponential and power-law cross section variations (\ref{profile}) had been discussed in the literature \cite{rud-aaa-2002}; as for integral representation (\ref{solution1}) of the solution to the MGWE, it has presumably been obtained for the first time in this paper. The invariant solution given by Eqs. (\ref{solutiongen}), (\ref{selfsimilar2}), and (\ref{selfsimilar3}) has never been reported in the literature.

\section{Approximate symmetry group and approximate group invariant solutions to the MGWE \label{approxanalytics}}

The invariant solutions to the MGWE obtained in the previous section have the following drawback: being exact, they can only be obtained for certain specific channel profiles and initial conditions, which are given by classifying relation (\ref{classify}) and the solution to equation (\ref{factoreq}). In this section, we present alternative solutions, namely, approximate analytic solutions to nonlinear boundary-value problem (\ref{webeq2}), and these solutions can be constructed for arbitrary initial conditions. An instrument for constructing such solutions is the approximate symmetry group, and the condition for the existence of the latter is the presence of the small parameter related to the relative slowness of variation in the cross-sectional area along the waveguide axis;
i.e., the smallness of the derivative ${{\rm d} (\ln \mu (\zeta)})/{\rm d} \zeta \equiv \mu_\zeta/\mu \ll 1$ (here, the subscript denotes the derivative with respect to the corresponding argument). In this case, the symmetry of the boundary-value problem under study is represented by a series expansion in powers of the small parameter, which allows us to obtain approximate analytic solutions to the problem with an acceptable accuracy.
\par

To construct an approximate analytic solution for a channel with a slowly varying cross section, we use the renormalization-group symmetry algorithm \cite{kov-ufn-2008} for boundary-value problem (\ref{webeq2}). According to perturbation theory, this algorithm allows us to extend the solutions in nonlinearity parameter $a$ to the region of finite values of this parameter. The general description of the algorithm can be found in \cite{kov-ufn-2008}, and the details of the corresponding calculations performed in the case of solving the boundary-value problem for the modified Burgers equation are given in \cite{kov-lga-94}.\par

To take into account the transformation of parameter $a$, we include the latter in the list of independent
variables and represent the infinitesimal operator of this transformation as
\begin{equation}
  \begin{aligned}
  X_5 = \xi(a) \left( \frac{\partial }{\partial{a}} - \frac{q}{a} \frac{\partial }{\partial{q}} \right) \,.
 \label{gena}
 \end{aligned}
 \end{equation}
 The desired renormalization-group symmetry operator is obtained as a linear combination of operator (\ref{gena}) with $\xi(a)=1$ and infinite subgroup operator (\ref{genweb2ext3}), which (as we have shown in the previous section), in the zero order in $\mu_\zeta/\mu$, is allowed by boundary-value problem (\ref{webeq2}):
\begin{equation}
 \begin{aligned}
  R =  \frac{\partial }{\partial{a}} + \left( k^{(0)}(\zeta,\tau,a) \exp \left( - \frac{a q}{\mu} \right) - \frac{q}{a} \right) \frac{\partial }{\partial{q}}  \,.
 \label{rg1}
\end{aligned}
\end{equation}
Here, the function of three variables $k^{(0)}(\zeta,\tau,a)$ obeys linear parabolic equation (\ref{genweb2ext3}) with the initial condition $k^{(0)}(0,\tau,a)=W(\tau)/a$ determined by invariance of the solution at $a \to 0$ with respect to the renormalization-group symmetry operator (\ref{rg1}). As a result, we arrive at the expression
\begin{equation}
  \begin{aligned} &
  k^{(0)} = \frac{\nu}{a} K_a \,, \ K(a,x,\tau)= \int\limits_{-\infty}^{\infty}
  {\rm d} \xi {\rm e}^{\frac{a W(\xi)}{\nu}} G(x,\tau-\xi)\,, \quad G(x,\tau)=
  \frac{1}{\sqrt{4 \pi \nu x}} {\rm e}^{-\frac{\tau^2}{4\nu x}} \,.
 \label{rg2}
 \end{aligned}
 \end{equation}
Here, the subscript marking function $K$ denotes the partial derivative with respect to the corresponding
argument, $K_a \equiv \partial K/\partial a$. \par

Finite transformations of the continuous group are related to the infinitesimal transformation in a one-to-one manner by the Lie equations, i.e., equations of the characteristics for the first-order partial differential equation conjugate to the operator (\ref{rg1}), (\ref{rg2}). The solution of the Lie equations for operator (\ref{rg1}), (\ref{rg2}) yields the following approximate analytic solution to initial problem (\ref{webeq2}):
\begin{equation}
  \begin{aligned}
  q^{(0)} = \frac{\mu}{a} \ln \left[  1+\frac{\nu}{\mu} \left( K -1 \right) \right]
  \,,
 \label{rgsol1}
 \end{aligned}
 \end{equation}
which is valid in a medium with a slowly varying cross section, $\mu_{\zeta}/\mu \ll 1$. In fact, the
derivation of solution (\ref{rgsol1}) from the solution obtained for a channel with a constant cross
section consists in the presence of a factor $\nu/\mu$ other than unity. \par

The advantage of the renormalization group method is the possibility of sequentially refining the resulting
analytic approximations. As applied to the problem under study, such a refinement (in the next,
i.e., first-order, approximation in $\mu_\zeta/\mu$) is achieved as follows: the function
$k^{(0)}(\zeta,\tau,a)$ in generator (\ref{rg1}) is replaced by $k^{(1)}(\zeta,\tau,a)$, for which, instead of using the solution to the parabolic equation (\ref{genweb2ext3}), we use the solution to the inhomogeneous parabolic equation
\begin{equation}
 \begin{aligned}
 \frac{\partial k^{(1)}}{\partial {\zeta}} - \mu \frac{\partial^2 k^{(1)}}{\partial {\tau}^2} = - A \,.
 \label{krg1}
 \end{aligned}
 \end{equation}
Here right-hand side is proportional to the gradient of the channel cross section $\mu_\zeta/\mu\ll1$ and
linearly depends on the function $q^{(0)}$ of the zero-order approximation in this gradient:
\begin{equation}
 \begin{aligned} &
  A = ak^{(0)}q^{(0)}\mu_{\zeta}/ \mu^2 = (\mu_{\zeta}/ \mu)(\nu/a) K_a \ln \left[1+(\nu/\mu)(K-1)
  \right]\,.
 \label{krg2}
 \end{aligned}
 \end{equation}
Equation (\ref{krg1}) is obtained at the stage of calculating the renormalization-group symmetry operator (\ref{rg1}) from the
so-called \textit{group determining equation}, where the contributions proportional to $\mu_\zeta/\mu$ (which were
omitted at the previous step) are calculated using the zero approximation results (\ref{rg2}) and (\ref{rgsol1}). The solution
to Eq. (\ref{krg1}) yields a modified (due to the contribution with the cross section gradient) expression for the
function $k^{(0)}(a,x,\tau) \Rightarrow {k}^{(1)}(a,x,\tau)$
\begin{equation}
  \begin{aligned}
  {k}^{(1)} & = \frac{\nu}{a} K_a - \int\limits_{0}^{x}{\rm d} x' \int\limits_{-\infty}^{\infty} {\rm d}\tau'
  G(x-x',\tau-\tau') \frac{\nu \mu'_{x'}}{a \mu'} K'_a \ln \left(1+ \frac{\nu}{\mu'}\left(K'-1\right) \right)
   \,, \\ & \quad \mu'\equiv \mu(x')\,, \quad K' \equiv K(a,x',\tau') \,.
 \label{rg3}
 \end{aligned}
 \end{equation}
The substitution of  $k^{(1)}$ instead of $k^{(0)}$ in infinitesimal operator (\ref{rg1}) and the subsequent solution of the Lie equations
leads to a refined approximation for the desired solution:
\begin{equation}
  \begin{aligned}
  q^{(1)}& = \frac{\mu}{a} \ln \left\{ 1 + \frac{\nu}{\mu} \left( K -1 \right)
  - \frac{\nu}{\mu}\int\limits_{0}^{x}{\rm d} x' \frac{\mu'_{x'}}{\mu'} \int\limits_{-\infty}^{\infty} {\rm d}\tau'
  G(x-x',\tau-\tau')  \right. \\ &
  \left. \times
   \left[ 1-K' + \left( K'-1+ \frac{\mu'}{\nu} \right)
  \ln \left(1+ \frac{\nu}{\mu'} \left( K' - 1  \right) \right) \right] \right\}
  \,.
 \label{rgsol2}
 \end{aligned}
 \end{equation}
For small values of the nonlinearity parameter  $a$, the first two terms of the expansion of solution (\ref{rgsol2}) in
powers of the nonlinearity parameter have the form
\begin{equation}
  \begin{aligned}
  q^{pt} & =  \nu K_a^{(0)} + \frac{\nu a}{2} \left[ \vphantom{\int\limits_{-\infty}^{\infty}} K_{aa}^{(0)} - \frac{\nu}{\mu} (K_a^{(0)})^2 \right.
  \left.
  - \int\limits_{0}^{x}{\rm d} x' \frac{\nu \mu'_{x'}}{(\mu')^2} \int\limits_{-\infty}^{\infty} {\rm d}\tau'
  G(x-x',\tau-\tau')  ((K')_a^{(0)})^2 \right] + O(a^2) \,,
 \label{rgsolpt}
 \end{aligned}
 \end{equation}
where $K_a^{(0)}$ and $K_a^{(0)}$ represent the values of the partial derivatives of function $K$ calculated in the limit a $a\to 0$.
By substituting the periodic initial condition $W(\xi)=\cos \xi$ in $K$ and calculating the resulting integrals, it is
possible to verify that the expression for $q^{pt}$ agrees well with the result obtained in \cite{rud-acj-1990} for a harmonic initial
perturbation.
\par

In closing this section, we present the form of the solution to problem (\ref{webeq2}) for a periodic initial condition  $W(\xi)=\cos \xi$: it is given by Eq. (\ref{rgsol2}) with function $K$ determined by the expression
\begin{equation}
  \begin{aligned}
  K = I_0(a/\nu) + 2\sum\limits_{k=1}^{\infty} I_k(a/\nu) \cos k\tau \, {\rm e}^{-\nu k^2 x} \,.
 \label{rgsolcos}
 \end{aligned}
 \end{equation}
The use of Eqs. (\ref{rgsol2}) and (\ref{rgsolcos}) allows us to calculate the nonlinear distortion of the spectrum of an acoustic wave propagating in a waveguide with a varying cross section. This opens up better possibilities for diagnostics of acoustic propagation paths, as compared to the weakly nonlinear limit \cite{rud-acj-1990}.
\begin{figure}[!h]
\centerline{\includegraphics[width=0.92\textwidth]{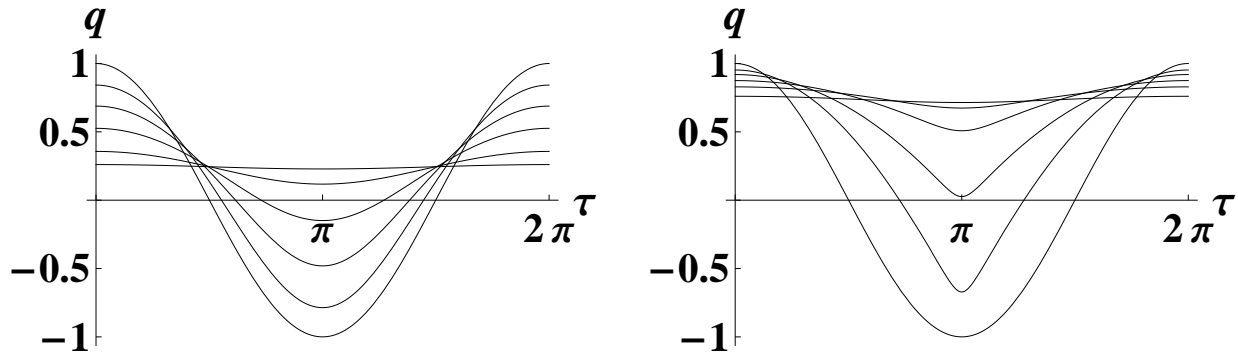}}
\caption{Curves representing the solution to modified generalized Webster equation (\ref{webeq2}) versus quantity $\tau$. The curves are obtained from Eqs. (\ref{rgsol2}) and (\ref{rgsolcos}) for the periodic initial condition $W(\xi)=\cos \xi$ for different values of the $x$ coordinate along the axis of the channel whose cross-sectional area varies according to the exponential law $\mu(x)/\nu=\exp(\alpha x)$. The left-hand plot shows the curves for $\nu x$ = 0, 0.2, 0.5, 1, 2, and 4 (the increase in $\nu x $ corresponds to the passage from the upper curves to the lower ones at the axis $\tau = 0$) and for $\alpha/\nu = -0.1$ and $a/\nu=1$. The right-hand plot shows the curves for $\nu x =$ 0, 0.08, 0.2, 0.5, 1, and 2 (the increase in $\nu x $ corresponds to the passage from the upper curves to the lower ones at the axis $\tau = 0$) and for $\alpha/\nu = -0.1$ and $a/\nu=10$.
\label{web110}}
\end{figure}
As an example, in Fig.~\ref{web110} we represent the solutions to the MGWE, $q \equiv q^{(1)}$, versus $\tau$, the solutions being obtained from Eqs. (\ref{rgsol2}) and (\ref{rgsolcos}) for different values of the $x$ coordinate along the axis of a channel with an exponentially varying cross-sectional area, $\mu/\nu=\exp(\alpha x)$. \par

\begin{figure}[!ht]
\centerline{\includegraphics[width=0.92\textwidth]{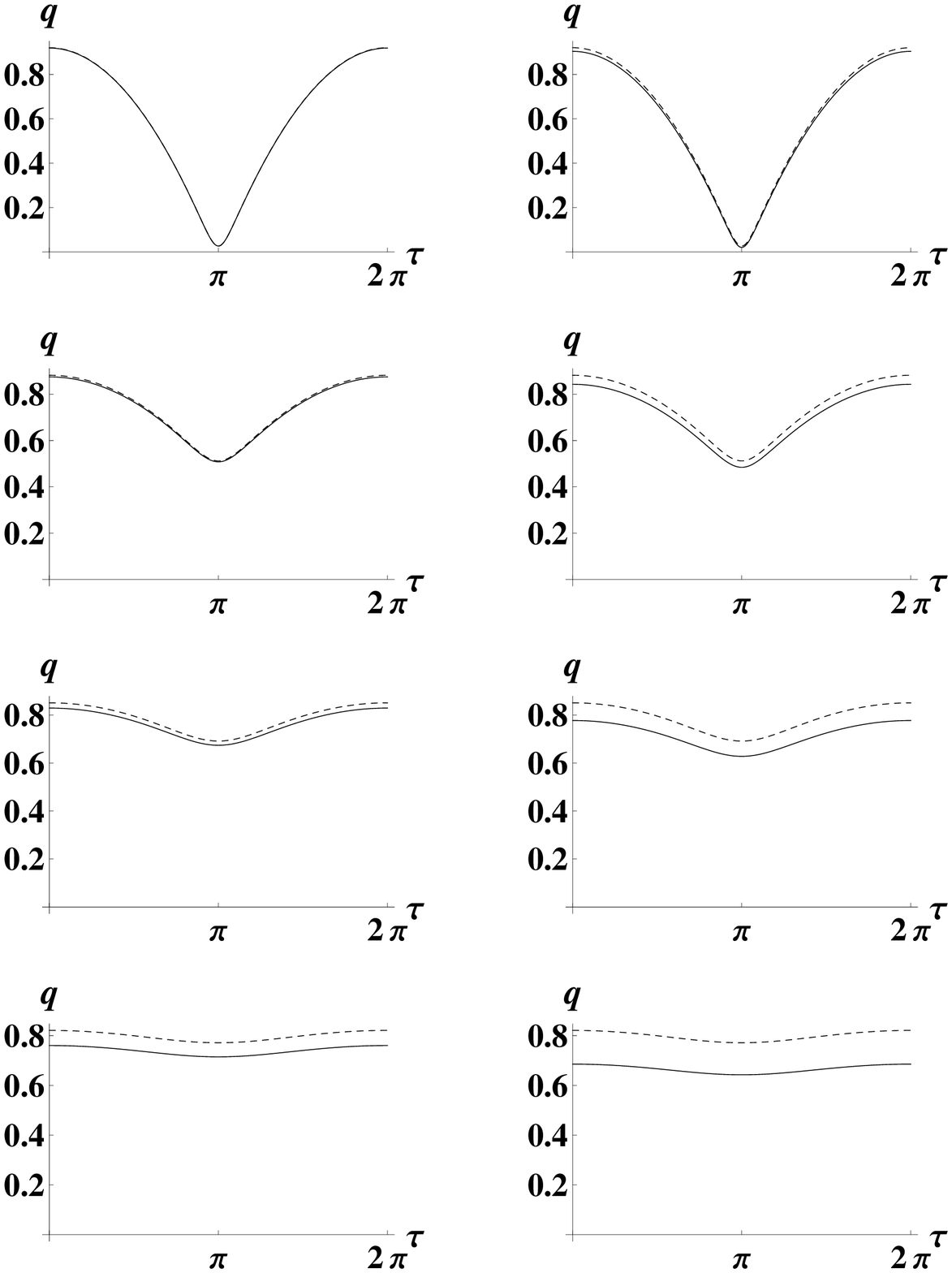}}
\caption{Comparison between obtained solutions $q$ to modified generalized Webster equation (\ref{webeq2}) by numerically solving Eq. (\ref{webeq2})
(dashed curves) and the approximate analytic solutions $q \equiv q^{(1)}$ (solid curves in the left-hand plot) and $q \equiv q^{(0)}$ (solid curves in the right-hand plot) for different values of the coordinate: $\nu x =$ 0.2, 0.5, 1, and 2. The increase in the $\nu
x$ coordinate along the pipe axis corresponds to the passage from the upper plots to the lower ones. The values of other parameters are the same as those in Fig.~\ref{web110}. \label{q1q0qnum}}
\end{figure}

To estimate the accuracy of the approximate analytic solutions obtained by us, we numerically solved initial equation (\ref{webeq2}). Comparison of the curves plotted with the use of analytic results (\ref{rgsol2}) and (\ref{rgsolcos}) with the curves obtained from the numerical solution of initial equation (\ref{webeq2}) shows good agreement between the numerical and analytic results for the case of a moderate nonlinearity $a/\nu = 1$ with an accuracy of up to fractions of percent. The results of calculating $q \equiv q^{(1)}$ for a greater nonlinearity $a/\nu = 10$ show a difference between the results of numerical and analytic calculations; the difference increases with increasing distance along the pipe, as one can see from the comparison of the curves shown in Fig.~\ref{q1q0qnum} at the left. However, the value of the difference is relatively small, and, even for $x=2$, it is on the order of seven percent. We also note that the strongest effect of nonlinearity, which makes the wave profile steeper, manifests itself as early as in the region of $x$ where the difference between numerical and analytic results is small: for the example under consideration with $a/\nu =10 $, the corresponding value of the $\nu x$ coordinate proves to be on the order of 0.08.
To illustrate the possibility of increasing the accuracy of analytic calculations with the use of approximate symmetry, in Fig.~\ref{q1q0qnum} (right), we present the curves showing the difference between the zero-approximation solution $q^{(0)}$ and the solution $q$ obtained from numerical simulation. One can see that the change from $q^{(0)}$ to $q^{(1)}$ already considerably improves the agreement between the numerical results and the approximate analytic ones.

\section{Conclusion}

Summarizing the results, we note that, in the present paper, we used the modified generalized Webster equation as the model to study exact and approximate analytic solutions to the problem of sound wave propagation in a channel with a varying cross section. The application of group methods allowed us to determine the specific types of cross section profiles for which the aforementioned problem allows new exact group invariant solutions. The application of renormalization-group symmetries allowed us to determine approximate analytic solutions for arbitrary initial conditions and sufficiently smooth profiles of variation in the waveguide cross section and to demonstrate the way of refining the approximate solutions. \par

In addition to the conventional applications discussed in the Introduction, it is necessary to point out the possibility of using the results of our study in topical interdisciplinary investigations, which are primarily related to studies of biological tissues and medical applications. For example, equations of the type of the modified and generalized Webster equations had been used in hemodynamics for describing the nonlinear pulse waves \cite{rud-acj-09}. Inhomogeneities of blood vessels are characterized by an inner cross-sectional area $S(x)$ that decreases with distance from the heart. In addition, the stiffness of the vessel channel varies. The stiffness, which affects the wave propagation velocity, is determined by the varying ratio between the collagen and elastin contents. Peripheral vessels are usually stiffer. The diagnostics of blood vessels is an important
application of mathematical models based on the Webster-type equations and their complicated modifications, which are to be studied in the future.

\smallskip
\par

We are grateful to Prof. D.~V.~Shirkov for stimulating discussions that were aimed at revealing classical analogs of dimensional reduction and variable geometry effects in wave propagation problems described by equations of mathematical physics.

\smallskip
\par

This work was carried out at the Ufa State Aviation Technical University and the Lobachevski State University, Nizhni Novgorod, under contract nos.~11.G34.31.0042 and 11.G34.31.0066 according to resolution no. 220 of the Russian Federation Government. We are also grateful for the support of the Russian Foundation for Basic Research (project nos.~11-01-00267-a and 12-01-00940-a) and the Presidential Program in Support of Leading Scientific Schools of Russia (grant no. NSh-3810.2010.2).


\end{document}